\documentclass[10.5pt,oneside,english]{article}
\usepackage[a4paper, total={7.0in, 9.5in}]{geometry}
\usepackage[utf8]{inputenc}
\pdfoutput=1
\usepackage{titlesec}
\titlespacing*{\section}{0pt}{0.1\baselineskip}{0.2\baselineskip}
\usepackage{rotating}
\usepackage{graphicx}
\usepackage{amssymb}
\usepackage[table]{xcolor}
\usepackage{xcolor}
\usepackage[T1]{fontenc}
\usepackage{lmodern}

\usepackage{abstract}
\usepackage{multirow,booktabs,array,longtable,caption}

\usepackage{pdflscape}

\definecolor{verylightgray}{gray}{0.9}

\usepackage{enumitem}
\newlist{compactimize}{itemize}{1}
\setlist[compactimize]{label=\textbullet, nosep, left=0pt,
                    before={\begin{minipage}[t]{\hsize}},
                    after ={\end{minipage}} }

\newcommand{\changess}[1]{#1}

\usepackage{caption}

\usepackage{courier}  

\usepackage[bottom, multiple]{footmisc}  
\usepackage[normalem]{ulem}    
\usepackage{anyfontsize}

\usepackage[square, sort, comma, numbers]{natbib}
\PassOptionsToPackage{hyphens}{url}\usepackage[hidelinks, hyperfootnotes=false]{hyperref}
\usepackage{fancyhdr}                           
\twocolumn
\fancypagestyle{firstpage} 
{
   \footskip=12mm
   \fancyhf{}
   \fancyfoot[C]{\vspace{-6mm}\thepage}

   \fancyfoot[L]{\vspace{-0mm}\noindent\rule{7.0in}{0.4pt}
     \scriptsize{
         \textcopyright \hspace{0.5mm} Barclays 2024 \\
       This work is licensed under a \href{https://creativecommons.org/licenses/by/4.0/}{Creative Commons Attribution 4.0 International License (CC
       BY).}  \\ Provided you adhere to the CC BY license, including as to attribution, you are
       free to copy and redistribute this work in any medium or format and remix, transform,
       and build upon the work for any purpose, even commercially.                  \\
       BARCLAYS is a registered trademark, all rights are reserved. \\
     }
 }
}
\usepackage{sectsty}
\allsectionsfont{\raggedright}

\title{\vspace{-2cm}Anchoring UK Retail Digital Money}

\author{%
  \begin{tabular}{c} {\fontsize{10.75}{-0.5cm}\selectfont Lee Braine, Shreepad Shukla and Piyush Agrawal} 
    \\ {\fontsize{10.75}{1cm}\selectfont Chief Technology Office, Barclays}
  \\ \hskip 1em \end{tabular} }

  \date{\vspace{-.7cm}September 27, 2024}

\begin{document}
\maketitle
\thispagestyle{firstpage} 
\vspace{-0.75cm}


\renewcommand{\abstractnamefont}{\normalfont\normalsize\bfseries} 
\renewcommand{\abstracttextfont}{\normalfont\normalsize}

\begin{abstract}
  \vspace{-2mm}
  \noindent
\textit{In the UK, the Bank of England and HM Treasury 
are exploring a potential
UK retail CBDC, the digital pound, with one of their motivations 
being the potential role of the digital pound as an anchor
for monetary and financial stability. 
In this paper, we explore 
three \changess{elements for anchoring money} (singleness of money, official currency 
as the unit of account, and safety and soundness of 
financial institutions and payment systems)
that maintain public trust and confidence in 
private UK retail digital money and the financial system. 
We also identify core capabilities
(comprising on-demand interoperability 
across issuers and forms of private money, settlement finality 
in wholesale central bank money, and access to physical cash) 
and appropriate measures (comprising customer funds protection, 
robust regulation, effective supervision, 
safe innovation in money and payments, and the central bank 
as the lender of last resort)
that together provide the foundations for 
the three elements for anchoring money.
Our preliminary analysis concludes that anchoring 
private UK retail digital money  
is supported by these elements, capabilities and measures.
Further work could include public-private 
collaboration to explore anchoring all forms
of UK retail digital money.
}
\end{abstract}

\vspace{-5mm}
\section{Introduction}\label{sec:introduction}
\vspace{-2mm}

\noindent
Physical cash is the only public money available 
to households and businesses in the UK at present. 
However, in recent years, the use of physical cash for payments 
has declined \cite{uk-payment-trends} and technology advancements have increased the likelihood 
that new forms 
of private digital money could emerge. 
The Bank of England considers that these trends could threaten the uniformity\footnote
{``Uniformity or singleness means that all forms of money - both public and private, 
bank deposits and cash - are valued equally (`at par' or `face value'), 
denominated in a common currency (sterling) and interchangeable with each other'' \cite{boe-cbdc-cons-pape-resp}.}
of money (also known as the singleness of money), and potentially fragment 
the monetary system, posing a risk to monetary and financial stability. 
The Bank of England identified its primary motivations 
for a potential UK retail CBDC, the digital pound, in the 
Consultation Paper (CP) \cite{boe-cbdc-cons-paper} as:
(i) sustained access to 
central bank money as an anchor for confidence and safety in 
the monetary system, and 
(ii) promotion of innovation, choice and efficiency in domestic 
payments.

In this paper, we first explore the concept of 
anchoring money. 
We then identify and analyse three \changess{elements for anchoring money} 
that maintain public trust and 
confidence in private UK retail digital money.\linebreak 
We next identify core capabilities and appropriate measures 
that provide the foundations for the three \changess{elements}.
Finally, we conclude and identify potential next steps.
The key contribution of this paper is a perspective on 
anchoring private UK retail digital money supported by 
three elements, three core capabilities and four measures.

\vspace{-3mm}
\section{Anchoring in the existing UK monetary system}\label{sec:anch-exist-money}
\begin{figure}[!h]
  \begin{center}
  \includegraphics[width=0.95\linewidth]{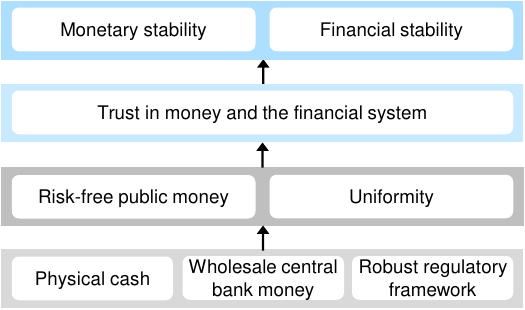}
  \end{center}
  \vspace{-6mm}
  \caption{\footnotesize{
    Three pillars of the monetary system that deliver uniformity and trust in
    money and the financial system. Figure adapted from \cite{boe-cbdc-cons-paper}.
  }}
 \label{fig:fig1-exist-anchor-model}

\end{figure}

\vspace{-2mm}
In the CP \cite{boe-cbdc-cons-paper}, the Bank of England identified 
three pillars of the existing monetary system 
(see Figure \ref{fig:fig1-exist-anchor-model}) that deliver uniformity,  
and trust in money and the financial system:
(i) physical cash, 
(ii) wholesale central bank money, and
(iii) robust regulation and supervision.
Central bank money (physical cash and reserves) is considered 
as an anchor for confidence and safety in the UK monetary 
system because it is close to risk-free \cite{boe-cash-risk-free}, 
can be elastically supplied, serves as 
a settlement asset and store of value, and provides the 
unit of account for the economy. 
Hence central bank money 
underpins monetary and financial stability, and sovereignty.

\vspace{-6mm}
\section{Related work on retail CBDC as a monetary anchor}\label{sec:anch-cbdc-global}
\vspace{-2mm}
In this section, we summarise related work that either 
supports retail CBDC as a monetary anchor or
discusses whether a retail CBDC is needed as a monetary anchor.

In the UK, given the decline in the share of payments 
using physical cash 
and the emergence of new forms of private digital money, the 
Bank of England believes there may be a future need for 
a digital pound to sustain retail access to central bank money and 
ensure its role as an anchor for confidence and 
safety in money \cite{boe-cbdc-cons-paper}. 

The European Central Bank believes
that, in future, a digital euro could help ensure 
central bank money remains a monetary anchor for the payments 
ecosystem and continues to serve as a means of exchange, a store of 
value and a unit of account \cite{ecb-cbdc-montryanchor}.

The Reserve Bank of New Zealand believes that a retail CBDC 
could directly support the role of central bank money as a monetary 
anchor by providing a digital alternative to privately 
issued money \cite{rbnz-cbdc}.

Rivadeneyra et al \cite{boc-role-pub-money} at the Bank of Canada suggested that a monetary system without a relevant 
form of retail public money would be at greater risk of losing monetary 
sovereignty and its financial regulatory regime may become less 
effective. 
They suggested that preserving continuity in the monetary 
system would likely require a retail CBDC to fulfill the role 
of cash in a digitalised economy, along with updates to the
financial regulatory regime.

Armelius et al \cite{riksbank-cbdc-anchor} at the Sveriges Riksbank and the Bank of Canada argued that ``instituted measures 
like deposit insurance, lender of last resort, regulations and supervision, 
together with sound government finances and macroeconomic policies make commercial 
bank money safe up to the limit of the deposit insurance guaranteed
and often beyond. Thus, to begin with, neither cash nor a CBDC seems fundamental 
to the monetary systems in countries with these measures in place'' (page 30).

Bofinger and Haas \cite{digeuro-anchor-suerf} at the University of W{\"u}rzburg identified
three potential anchor roles for a retail CBDC:
(i) anchor for commercial bank deposits, 
guaranteeing their convertibility into public money,
(ii) anchor to maintain the national currency 
as a unit of account, and 
(iii) anchor for maintaining the central bank's 
control over the financial system.
For the three potential anchor roles, they argued that: 
(i) a digital euro can be justified as an anchor for bank 
deposits, but this would require unlimited access to a store of value, 
whereas the ECB envisages very limited digital euro holdings as a means of payment,
(ii) anchoring the national currency as a unit of account requires stability 
of the currency and not the use of digital euro in daily payments, and
(iii) for the central bank's control over the financial system, it is crucial 
that banks use wholesale central bank money as a means of settlement.
They concluded that ``a digital euro is not needed as a monetary anchor for the euro area'' (page 7).

\vspace{-3mm}
\section{Examining the role of a digital pound in anchoring money}\label{sec:anch-money}
\vspace{-2mm}
In this section, we analyse the concept of anchoring money, 
including the role of a potential digital pound and other
mechanisms to anchor UK retail digital money.

We assume that \textit{\textbf{the primary aim of anchoring money is to 
maintain public trust and confidence in money (regardless of 
    its issuer and form) and the financial system}}.
    
    We identify and analyse the following three \changess{elements for anchoring money}:

\begin{enumerate}[label=(\roman*)]
  \item
    \textbf{Singleness of money:} 
    The CP \cite{boe-cbdc-cons-paper} stated that 
    ``stability of the UK economy and monetary
    system relies on the uniformity of money'' (page 25).
    Panetta \cite{ecb-cbdc-montryanchor} stated that convertibility into central bank money is 
    necessary for confidence in private money as a means of payment 
    and store of value. 

    However, cash and a potential digital pound are not the 
    only mechanisms to convert private money at par. 
    In the UK, the Bank of England operates the Real-Time Gross Settlement 
    (RTGS) system \cite{boe-rtgs}
    that allows firms to settle their payment obligations
    in wholesale central bank money.  
    This guarantees settlement finality of payments at par  
    in sterling.
    The Bank of England states that wholesale central bank money 
    ``therefore plays an anchoring role for commercial bank money'' \cite{boe-dp} (page 23).
    Garratt and Shin \cite{bis-singleness-money} suggested that settlement using central bank money 
    is the key feature that promotes singleness of money. 
    In the UK, in 2023, commercial bank money amounted to approximately 
    \pounds3 trillion and central bank money amounted to approximately
    \pounds964 billion 
    (\pounds87 billion of physical banknotes and \pounds878 billion of 
    wholesale money) \cite{boe-ab-singleness-money}. 
    High holding limits for retail CBDC may be considered if there was 
    an aim to \textit{fully} anchor the singleness of money.
    However, during stressed financial market conditions, a retail CBDC would 
    increase the likelihood and scale of commercial bank deposit outflows, which would
    ``increase the vulnerability of the banking sector to stress,
    potentially further destabilising credit supply and could require
    the central bank to act as a lender of last resort more quickly than it
    otherwise would need to''
    \cite{uk-finance-cbdc-fin-Stability} (page 17). 

    Singleness could therefore potentially be achieved with: 
    (i) on-demand interoperability
    between issuers and forms of private money
    (e.g. see interoperability design options in \cite{barc-usecase}), \linebreak
    (ii) settlement finality in wholesale central bank money, and
    (iii) prudent regulations (including a compensation scheme) and supervisions by the 
    Bank of England and the UK Government. 

  \item 
    \textbf{Official currency as the unit of account:}
      The role of an official currency as a unit of account is established and 
      maintained by central bank money, particularly in the form 
      of central bank reserves. 
      Central bank rates and quantitative easing are the two main monetary policy tools
      used by the Bank of England to influence the rate of inflation in the UK \cite{boe-mon-policy}.
      This underpins confidence in the value of money and provides  
      consumers and businesses with confidence to continue to use sterling 
      as the unit of account in the UK.

      As previously mentioned, the decline in the use of cash in 
      UK retail transactions is one of the Bank of England's motivations for the  
      introduction of a potential digital pound. 
      However, this reduction in the use of cash does not impact the 
      Bank of England's ability to control inflation because central 
      bank rates and quantitative easing
      are applied using whole central bank money denominated in sterling.
      
      If a new form of non-sterling digital money were to 
      become widely used in the UK, this could compromise monetary 
      sovereignty (i.e. the UK authorities' ability to achieve price stability 
      through monetary policy).
      The Bank of England envisages that innovative functionalities in the 
      digital pound could support maintaining sterling as the unit of account \cite{boe-cbdc-cons-paper}. 

      However, to avoid the risks of fragmentation in payments 
      markets, we emphasise that  
      innovative functionalities should be provided across all forms of 
      digital money, and not only on the digital pound \cite{barc-cbdc-iia}. 
      Our previous paper \cite{barc-cbdc-fc} explored the important concept
      of functional consistency across all forms of regulated retail digital money,
      and presents design options to support functional consistency and 
      interoperability across the digital pound and commercial bank money.
      Commercial bank money is deeply integrated in the UK economy and is well-positioned 
      to provide the innovative capabilities (such as programmable payments) in the near 
      future.

      Maintaining sterling as the unit of account could therefore be supported by:  
      (i) implementing measures to regulate access and use of new forms of 
      money in the UK, and
      (ii) providing innovative capabilities on commercial bank money.

  \item 
    \textbf{Safety and soundness of financial institutions
      and payment systems:}
      Central banks play a pivotal role in maintaining the safety and integrity 
      of payment systems, and provide a solid foundation by acting as 
      guardians of the stability of money and payments \cite{bis-cb-pay-dig-era}.
      In the UK, the Bank of England provides stability to the financial system 
      by regulating and supervising commercial banks and other financial institutions. 
      The Bank of England also promotes safety, efficiency and integrity of 
      payment systems by operating core infrastructure 
      (such as CHAPS \cite{boe-rtgs} and RTGS) 
      and overseeing other payment systems (such as Faster Payment System \cite{abt-fps}).

      Safety and soundness of financial institutions and payment systems 
      could therefore be achieved with:
      (i) the central bank as the lender of last resort,
      (ii) robust regulations and prudent policies 
      (such as Capital Requirements Directive \cite{abt-crd} 
      and Deposit Guarantee Scheme \cite{boe-dgs}), and
      (iii) effective supervision of commercial banks and payment systems. 
      These measures support the value of deposits
      (for example, by providing quick and orderly resolution 
      in the event of commercial bank or building society failure)
      and the availability and efficiency of payment systems.

\end{enumerate}

\vspace{-6mm}
\section{Foundations for anchoring retail private digital money}\label{sec:anch-all-money}
\begin{figure}[!h]
  \begin{center}
  \includegraphics[width=1\linewidth]{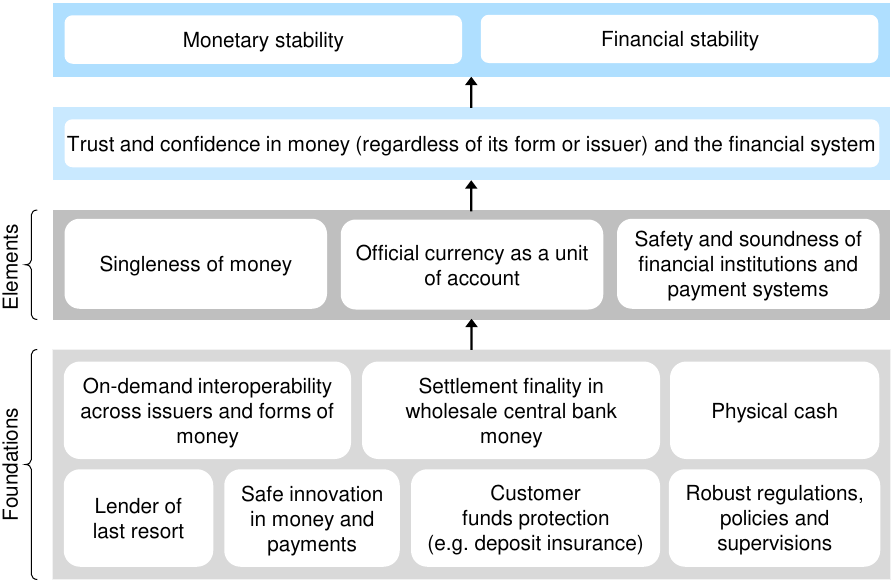}
  \end{center}
  \vspace{-6mm}
  \caption{\footnotesize{Anchoring private UK retail digital money.
  }}\label{fig:fig2-options-anchor-money}
\end{figure}

\vspace{-2mm}
Maintaining safety and trust in private UK retail digital money requires
strong foundations which ensure that
financial institutions are sound, payment systems are robust and 
innovation is enabled.
In this section, we identify core capabilities and appropriate
measures that support the three elements 
(analysed in previous section) for anchoring private 
UK retail digital money.
Figure \ref{fig:fig2-options-anchor-money} presents 
the three elements and important foundations\footnote{
  Further analysis could explore the degree to which each of these foundations
  support anchoring private UK retail digital money.
}, comprising core capabilities and appropriate measures,
that support trust and confidence in private UK retail digital money 
and the financial system. 
\changess{The core capabilities comprise on-demand interoperability 
across issuers and forms 
of private money, settlement finality in wholesale central bank money,
and access to physical cash. }
The appropriate measures comprise customer funds protection (e.g. deposit
insurance), robust regulation (e.g. capital requirements), effective supervision, 
safe innovation in money and payments, and the central bank as the lender of last resort.

\vspace{-2mm}
The Bank of England is modernising the wholesale payments infrastructure and
has introduced a new omnibus account model \cite{boe-omnibus}
in the current RTGS system, 
on which the private sector can provide innovative services.
Commercial banks are exploring payments innovations, such as 
programmable payments and tokenised deposits
(e.g. the UK Regulated Liability Network experimentation phase \cite{uk-rln}).

\section{Summary and conclusions}\label{sec:conclusions}
\vspace{-2mm}
In this paper, we explored the important concept of anchoring money, and
identified and analysed three \changess{elements for anchoring money}  
(i.e. singleness of money, official currency 
as the unit of account, and safety and soundness of 
financial institutions and payment systems)
that maintain public trust and confidence in  
private UK retail digital money and the financial system. 
We also identified core capabilities
(comprising on-demand interoperability 
across issuers and forms of private money, settlement finality 
in wholesale central bank money, and access to physical cash) 
and appropriate measures (comprising customer funds protection, 
robust regulation and effective supervision, 
safe innovation in money and payments, and the central bank 
as the lender of last resort)
that together provide the foundations for 
the three elements for anchoring money.

Our preliminary analysis concluded that the \linebreak
anchoring of private UK retail digital money  
is supported by the elements, core capabilities and appropriate 
measures identified above.
These could be sufficient to maintain public trust and confidence
in money and the financial system, despite a decline in 
the use of physical cash for payments and the emergence of new forms of private digital money.
However, further research is required before drawing definite conclusions.

\vspace{-4mm}
\section{Further work}\label{subsec:further-work}
\vspace{-2mm}
Potential further work could include public-private 
collaboration to explore options to anchor all forms 
of private UK retail digital money (including commercial bank money, 
e-money and regulated stablecoins).
This work should consider both normal and systemic stress 
conditions.

We hope this paper will inspire industry
collaboration on anchoring trust and confidence in money and the financial system, 
and we look forward to feedback.

\vspace{1mm}
\noindent \textbf{Acknowledgements:} 
\noindent
We thank Ekaterina Marshall (Barclays) 
for helpful feedback on this paper.

\vspace{-4mm}
\pretolerance=-1
\tolerance=-1
\emergencystretch=0pt

\bibliography{uk-cbdc-anchor-position-paper}

\begin{thebibliography}{10}

\bibitem{riksbank-cbdc-anchor}
Hanna Armelius, Carl~Andreas Claussen, and Scott Hendry.
\newblock
  \href{https://www.riksbank.se/globalassets/media/rapporter/pov/artiklar/engelska/2020/200618/2020_2-is-central-bank-currency-fundamental-to-the-monetary-system.pdf}{Is
  central bank currency fundamental to the monetary system?}
\newblock {Working Paper}, {Sveriges Riksbank}, May 2020.

\bibitem{boe-mon-policy}
{Bank of England}.
\newblock \href{https://www.bankofengland.co.uk/monetary-policy}{Monetary
  policy}.
\newblock Accessed September 2024.

\bibitem{boe-rtgs}
{Bank of England}.
\newblock
  \href{https://www.bankofengland.co.uk/payment-and-settlement/a-brief-introduction-to-the-real-time-gross-settlement-system-and-chaps}{A
  brief introduction to the Real-Time Gross Settlement system and CHAPS}.
\newblock Accessed September 2024.

\bibitem{abt-crd}
{Bank of England}.
\newblock
  \href{https://www.bankofengland.co.uk/prudential-regulation/key-initiatives/capital-requirements-directive-iv}{Capital
  Requirements Directive}.
\newblock Accessed September 2024.

\bibitem{boe-cash-risk-free}
{Bank of England}.
\newblock
  \href{https://www.bankofengland.co.uk/-/media/boe/files/quarterly-bulletin/2014/the-boe-as-a-bank.pdf}{The
  Bank of England as a bank}, June 2014.

\bibitem{boe-omnibus}
{Bank of England}.
\newblock
  \href{https://www.bankofengland.co.uk/news/2021/april/boe-publishes-policy-for-omnibus-accounts-in-rtgs}{Bank
  of England publishes policy for omnibus accounts in RTGS}, April 2021.

\bibitem{boe-dgs}
{Bank of England}.
\newblock
  \href{https://www.bankofengland.co.uk/-/media/boe/files/prudential-regulation/statement-of-policy/2023/deposit-guarantee-scheme-june-2023.pdf}{Deposit
  Guarantee Scheme}.
\newblock Statement of policy, June 2023.

\bibitem{boe-ab-singleness-money}
{Bank of England}.
\newblock
  \href{https://www.bankofengland.co.uk/-/media/boe/files/speech/2023/july/new-prospects-for-money-speech-by-andrew-bailey}{New
  prospects for money - speech by Andrew Bailey}.
\newblock July 2023.

\bibitem{boe-dp}
{Bank of England}.
\newblock
  \href{https://www.bankofengland.co.uk/-/media/boe/files/paper/2024/the-boes-approach-to-innovation-in-money-and-payments.pdf}{The
  Bank of England's approach to innovation in money and payments}.
\newblock {Discussion Paper}, July 2024.

\bibitem{boe-cbdc-cons-paper}
{Bank of England and HM Treasury}.
\newblock
  \href{https://www.bankofengland.co.uk/-/media/boe/files/paper/2023/the-digital-pound-consultation-working-paper.pdf}{The
  digital pound: A new form of money for households and businesses?}
\newblock {Consultation Paper}, February 2023.

\bibitem{boe-cbdc-cons-pape-resp}
{Bank of England and HM Treasury}.
\newblock
  \href{https://www.bankofengland.co.uk/-/media/boe/files/paper/2024/responses-to-the-digital-pound-consultation-paper.pdf}{Response
  to the Bank of England and HM Treasury Consultation Paper: The digital pound:
  a new form of money for households and businesses?}
\newblock {Consultation Response}, January 2024.

\bibitem{bis-cb-pay-dig-era}
BIS.
\newblock \href{https://www.bis.org/publ/arpdf/ar2020e3.pdf}{Central banks and
  payments in the digital era}.
\newblock {BIS Annual Economic Report}, {BIS}, June 2020.

\bibitem{digeuro-anchor-suerf}
Peter Bofinger and Thomas Haas.
\newblock
  \href{https://www.suerf.org/wp-content/uploads/2023/12/f_39cbf7bd666de7d755c556b218a29054_65263_suerf.pdf}{The
  Digital Euro (CBDC) as a Monetary Anchor of the Financial System}.
\newblock {Whitepaper}, University W{\"u}rzburg, April 2023.

\bibitem{barc-cbdc-iia}
Lee Braine and Shreepad Shukla.
\newblock
  \href{https://www.risk.net/journal-of-financial-market-infrastructures/7957848/illustrative-industry-architecture-to-mitigate-potential-fragmentation-across-a-central-bank-digital-currency-and-commercial-bank-money}{An
  Illustrative Industry Architecture to Mitigate Potential Fragmentation across
  Central Bank Digital Currency and Commercial Bank Money}.
\newblock {\em Journal of Financial Market Infrastructures}, 10(4):55--63,
  2022.

\bibitem{barc-cbdc-fc}
Lee Braine, Shreepad Shukla, and Piyush Agrawal.
\newblock \href{https://arxiv.org/pdf/2308.08362.pdf}{Functional Consistency
  across Retail Central Bank Digital Currency and Commercial Bank Money}.
\newblock August 2023.

\bibitem{barc-usecase}
Lee Braine, Shreepad Shukla, and Piyush Agrawal.
\newblock \href{https://arxiv.org/pdf/2409.08653}{Payments Use Cases and Design
  Options for Interoperability and Funds Locking across Digital Pounds and
  Commercial Bank Money}.
\newblock September 2024.

\bibitem{bis-singleness-money}
Rodney Garratt and Hyun~Song Shin.
\newblock \href{https://www.bis.org/publ/bisbull73.pdf}{Stablecoins versus
  tokenised deposits: implications for the singleness of money}.
\newblock {BIS Bulletin}~73, {BIS}, April 2023.

\bibitem{ecb-cbdc-montryanchor}
{Panetta, Fabio}.
\newblock
  \href{https://www.ecb.europa.eu/press/key/date/2021/html/ecb.sp211105~08781cb638.en.html}{Central
  bank digital currencies: a monetary anchor for digital innovation}.
\newblock November 2021.
\newblock Speech by Fabio Panetta, Member of the Executive Board of the ECB.

\bibitem{abt-fps}
{Pay.UK}.
\newblock
  \href{https://www.wearepay.uk/what-we-do/payment-systems/faster-payment-system/}{Faster
  Payment System}.
\newblock Accessed September 2024.

\bibitem{rbnz-cbdc}
{Reserve Bank of New Zealand}.
\newblock
  \href{https://www.rbnz.govt.nz/-/media/project/sites/rbnz/files/consultations/banks/future-of-money/cbdc-issues-paper.pdf}{The
  Future of Money - Central Bank Digital Currency}, Decemeber 2021.

\bibitem{boc-role-pub-money}
{Rivadeneyra, Francisco and Hendry, Scott and García, Alejandro}.
\newblock
  \href{https://www.bankofcanada.ca/wp-content/uploads/2024/07/sdp2024-11.pdf}{The
  Role of Public Money in the Digital Age}, July 2024.

\bibitem{uk-finance-cbdc-fin-Stability}
{UK Finance}.
\newblock
  \href{https://www.ukfinance.org.uk/system/files/2022-11/Retail%20UK%20CBDC%20on%20credit%20creation%20and%20financial%20stability.pdf}{Investigating
  the Potential Impact of a Retail UK CBDC On Credit Creation and Financial
  Stability}, November 2022.

\bibitem{uk-payment-trends}
{UK Finance}.
\newblock
  \href{https://www.ukfinance.org.uk/system/files/2024-07/Summary%20UK%20Payment%20Markets%202024.pdf}{UK
  Payment Markets Summary}, July 2024.

\bibitem{uk-rln}
{UK Finance}.
\newblock
  \href{https://www.ukfinance.org.uk/system/files/2024-09/UK%20Finance%20RLN%20Summary%20Report.pdf}{UK
  RLN Experimentation Phase - Summary Report}, September 2024.

\end{thebibliography}
\bibliographystyle{plain}
\end{document}